\documentclass{vgtc}                          

\ifpdf
  \pdfoutput=1\relax                   
  \usepackage{graphicx}                
  \DeclareGraphicsExtensions{.pdf,.png,.jpg,.jpeg} 
\else
  \usepackage{graphicx}                
  \DeclareGraphicsExtensions{.eps}     
\fi%

\graphicspath{{figures/}{pictures/}{images/}{./}} 

\usepackage{microtype}                 
\PassOptionsToPackage{warn}{textcomp}  
\usepackage{textcomp}                  
\usepackage{mathptmx}                  
\usepackage{times}                     
\usepackage{cite}                      
\usepackage{tabu}                      
\usepackage{booktabs}                  
\usepackage{multirow}
\usepackage{enumitem}
\usepackage{array}
\usepackage{balance}

\usepackage[table]{xcolor} 

\onlineid{0}

\vgtccategory{Research}

\vgtcinsertpkg

\title{Narrative Sensemaking: Strategies for Narrative Maps Construction}

\author{Brian Felipe Keith Norambuena\thanks{e-mail: briankeithn@vt.edu}\\ %
        \parbox{1.5in}{\scriptsize \centering Virginia Tech \\ Universidad Católica del Norte}%
\and Tanushree Mitra\thanks{e-mail: tmitra@uw.edu}\\ %
     \scriptsize University of Washington %
\and Chris North\thanks{e-mail: north@vt.edu}\\ %
     \scriptsize  Virginia Tech}

\abstract{Narrative sensemaking is a fundamental process to understand sequential information. Narrative maps are a visual representation framework that can aid analysts in this process. They allow analysts to understand the big picture of a narrative, uncover new relationships between events, and model connections between storylines. As a sensemaking tool, narrative maps have applications in intelligence analysis, misinformation modeling, and computational journalism. In this work, we seek to understand how analysts construct narrative maps in order to improve narrative map representation and extraction methods. We perform an experiment with a data set of news articles. Our main contribution is an analysis of how analysts construct narrative maps. The insights extracted from our study can be used to design narrative map visualizations, extraction algorithms, and visual analytics tools to support the sensemaking process.%
} 

\CCScatlist{
  \CCScatTwelve{Human-centered computing}{Visu\-al\-iza\-tion}{Empirical studies in visu\-al\-iza\-tion}{};
  \CCScatTwelve{Human-centered computing}{Visu\-al\-iza\-tion}{Visu\-al\-iza\-tion techniques}{Graph drawings}
}

\usepackage{fancyhdr}
\begin{document}



\maketitle\thispagestyle{fancy}

\section{Introduction} 
\label{sec:intro}
Narratives are systems of stories \cite{halverson2011master}---sequences of events tied together in a coherent fashion. Events are the fundamental units of narrative action, they are either an act involving characters and entities or a happening where no entities are causally involved \cite{abbott2008cambridge}. Narratives are fundamental to our understanding of the world and provide a natural way to capture relationships between sequences of events, as well as the goals, motivations, and plans of actors \cite{finlayson2013military}. Narratives are used in the process of ``connecting the dots" between apparently unrelated pieces of information \cite{hossain2011helping, hossain2012connecting}.

Storytelling in general is an accepted metaphor used in visual analytics and analytical reasoning \cite{cook2005illuminating,segel2010narrative,tong2018storytelling}. However, unlike general visual storytelling, our work focuses specifically on visualizing textual narratives, such as those created by news. In this context, narratives provide a way to understand the information landscape, a key part of several narrative sensemaking tasks \cite{keith2020maps}. For example, aiding in tasks such as journalistic analysis of news narratives \cite{bradel2015big} and intelligence analysis \cite{endert2014human}. To aid analysts with sensemaking tasks, scholars have created visual analytics software, which allow analysts to process and understand greater quantities of data and information \cite{cook2005illuminating}. These tools focus on different parts of the sensemaking loop \cite{pirolli2005sensemaking}, such as foraging \cite{kang2009evaluating} and synthesis \cite{wright2006sandbox}.

In this work, we focus on a specific type of graph-based narrative representation---narrative maps \cite{keith2020maps}. Narrative maps provide a generic foundation to encode any narrative extracted from data, requiring only the existence of a total ordering (e.g., timestamps) and text representation of the event (e.g., news headlines). Narrative maps are a useful visualization framework to understand the information landscape. As a sensemaking tool, they have applications in intelligence analysis, misinformation modeling, and computational journalism \cite{keith2020maps}. In particular, they offer a way to keep track of the big picture of a narrative in the context of the ever-increasing problem of information overload \cite{ho2001towards,shahaf2010connecting}. However, from a visualization standpoint, the construction process of narrative maps for sensemaking remains unexplored. Thus, our goal is to understand how analysts construct narrative maps to design better narrative representation and extraction methods. We focus on the strategies, cognitive connections, and structures used in the construction process.

\section{Related Work}
\label{sec:rel-work}
Narratives are systems of stories with coherent themes \cite{halverson2011master}. Narrative studies attempt to understand the relationships between the underlying stories and their representations \cite{abbott2008cambridge, puckett2016narrative}. In the context of information visualization, we explore how information narratives can be visualized. The narrative visualization pipeline requires extracting the narrative from data into an internal representation, which is then used to generate the final visualization \cite{keith2020maps}.

Storytelling and narratives are common metaphors in visual analytics \cite{tong2018storytelling}. In general, scholars have studied how arranging visualizations as story sequences helps in sensemaking \cite{hullman2013deeper,hullman2017finding}. Other works focused on narrative visualization for news usually focus on augmenting data visualization techniques (e.g., charts) with contextual information (e.g., relevant articles associated with data points in the chart) \cite{hullman2013contextifier, gao2014newsviews}. However, in our application context, we are interested in extracting and representing narratives taken directly from data sets of text documents, rather than augmenting numerical (or other non-text types of data) visualizations with contextual information or using sequences of visualizations to represent a story.

Previous research has explored how analysts make cognitive connections between documents in the context of intelligence analysis tasks. Bradel et al. \cite{bradel2013analysts} studied how analysts structure information, finding layouts based on linear structures with branching and web-like structures. Robinson \cite{robinson2008collaborative} focused on analyzing the strategies and organizational methods used in collaborative synthesis, with the purpose of proposing design guidelines for collaborative sensemaking systems. Andrews et al. \cite{andrews2010space} explored the workspace organization used by analysts in large displays to arrange documents, where most strategies consisted of clustering, although some analysts used timelines. Wenskovitch and North \cite{wenskovitch2020examination} studied how analysts perform grouping and dimensionality reduction, where strategies included divide-and-conquer, incremental layouts, and bottom-up construction. Our work follows a similar approach, but focusing exclusively on sensemaking with narrative maps, analyzing the different map construction strategies and structures generated in the process. 

Previous studies have found that analysts use strategies such as identifying co-occurrence relationships and aggregating common elements \cite{haider2017analysts}, using topical and temporal orderings for document clustering, and evaluating content overlap and similarity for document summarization \cite{endert2012clustering, camargo2013manual}. However, previous research has not focused on narratives, which have an underlying temporal ordering as well as a focus on cause-effect relationships that leads to a specific description of cognitive connections and construction strategies.

Finally, prior works have shown that graph-based narrative representations \cite{shahaf2013metro,keith2020maps,liu2017growing} are useful as a sensemaking tool. Thus, with the purpose of improving such narrative representations and their extraction algorithms, we seek to understand how analysts create such models from scratch by analyzing the narrative mapping process.

\section{Study Description}
\subsection{Data Set}
We used a data set comprised of 40 COVID-19 news articles from January 2020 covering the start of the Coronavirus outbreak. This data set is a subset of the COVID-19 archive data used in previous works on narrative maps \cite{keith2020maps}. The events were carefully curated in order to have a sufficiently small data set while covering a series of different topics and issues regarding the COVID-19 narrative. In particular, the articles cover topics such as the economic consequences of the pandemic, the sociopolitical effects in China, and the worldwide response. As our data set was made up of breaking news, the main event is usually described explicitly in the headlines \cite{norambuenaevaluating}. Thus, we focused on the headlines rather than the full article. We also included the publication dates and sources.  

\subsection{Task Definitions}
We defined two tasks to explore how analysts constructed narrative maps, a directed task that required participants to join two events and an open-ended task that required participants to expand on the outcomes of an initial event. In both tasks, participants were given a list of events (i.e., nodes) and asked to construct a narrative map by designing its overall structure, layout, and specific connections. The participants were also asked to label their main storyline---the core events of the narrative---and their side stories---stories relevant to the overall narrative but not directly related to the main storyline.

The directed task required participants to construct a narrative map to answer the following question: ''\textit{How did the Wuhan outbreak lead to the US travel restrictions?}", which referred to two specific events in the data set. This task is also known as ``connecting the dots" and it is a fundamental task in narrative sensemaking \cite{keith2020maps}. Previous research has attempted to understand how analysts perform this process \cite{bradel2013analysts} and sought to automate this process through algorithmic approaches \cite{shahaf2010connecting}. In contrast, the open-ended required participants to construct a narrative map to answer the following question: ''\textit{What outcomes occurred as a result of the Wuhan outbreak?}". This task is a variation of the ``connect the dots" task \cite{shahaf2012connecting} that only provides the starting event as a fixed point, requiring participants to explore the stories that emerge because of this event. 

Finally, both tasks required participants to label their storylines and to answer a follow-up question with their map: ``\textit{What are the key events (i.e., the most important events or turning points)?}". The focus of this experiment was to glean insights on the construction process, rather than comparing how the tasks themselves influence the construction. By considering two tasks rather than a single one, we expected to gather additional insights on this process.

\subsection{Evaluation Procedure}
\begin{figure*}[!htb]
    \centering
    \includegraphics[width=0.88\textwidth]{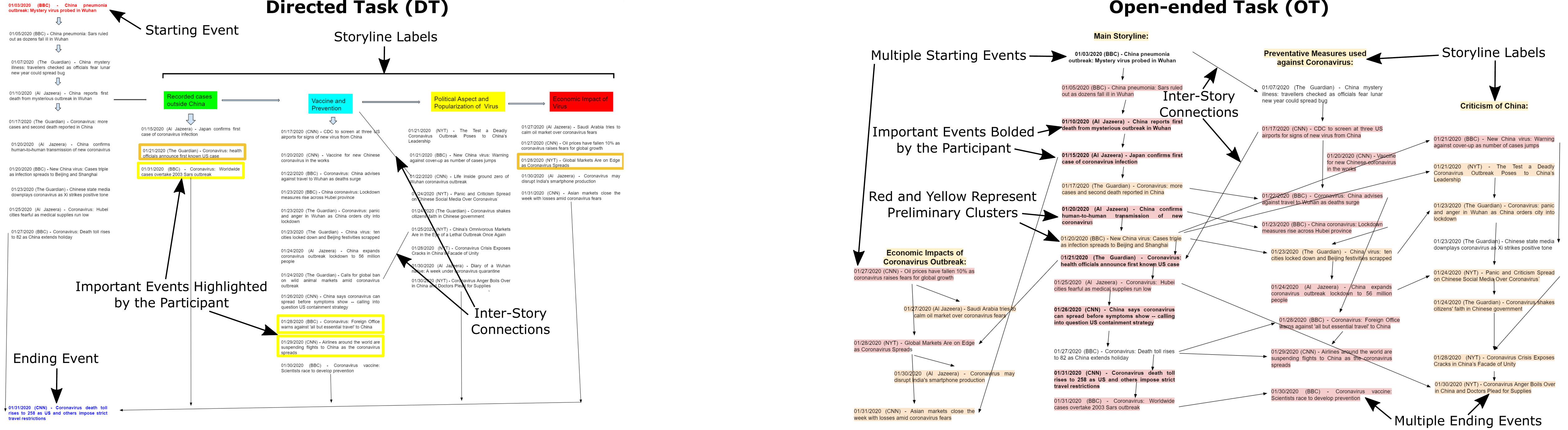}
     \caption{Narrative map examples created by participants for the two tasks \textbf{Directed Task} (DT) and \textbf{Open-ended Task} (OT). Annotations highlight key elements in these maps. Note how in the DT map all storylines converge into a single ending event, emphasizing the focused nature of this task. In contrast, the OT map has a series of storylines that interact with one another, representing the different outcomes found by the participant.}
    \label{fig:examples-rq1}
    \vspace{-12pt}
\end{figure*}

We recruited ten participants, following a similar approach to the work of Bradel et al. \cite{bradel2013analysts}. We assigned five participants to each task. While splitting the participants into two tasks increases variability, we expected to gather a wider range of construction strategies by doing this. All participants were part of a national security program and hence, had a background in intelligence analysis. They also had previous knowledge on the topic which they could leverage while conducting the tasks. Prior knowledge ranged from general knowledge about COVID-19 to stronger backgrounds since some participants were ardent followers of the pandemic news right from its start. Figure \ref{fig:examples-rq1} shows examples of the maps created in each task.

To provide initial training and to avoid inducing biases in subsequent task performance, participants were provided with a short example narrative map on a different topic. We engaged with our participants in an hour-long semi-structured session in a video call where they completed their assigned narrative sensemaking task. We encouraged the participants to think aloud and ask questions and share any observations as they worked. All participants were recorded and the videos were analyzed to understand their construction strategies. We explained that there were no correct or incorrect answers; as our goal was to understand the cognitive strategies used by the analysts to complete the tasks. However, the quality and conceptual cohesion varied among the solutions.

To construct the map, we gave participants a canvas on Google Drawings with the instructions and the list of articles chronologically ordered. The participants had to drag and drop the articles into the available space. Then, they had to add connections with arrows. The participants were instructed to design the map with other analysts as potential users in mind. The participants were familiar with Google Drawings and similar editing tools, thus they did not require additional training in its use, even if it might not have been their preferred tool for such an exercise. Moreover, they had full access to this tool through their institutional accounts. 

We opted for Google Drawings in our study for several reasons. First, it provided a closer approximation of what a computational narrative map tool would look like compared to an approach using hand-drawn notes. Thus, even though it might influence the kinds of strategies used by the participants, these strategies should be closer to what we would expect with a computational tool. Second, given the limitations caused by the pandemic, using Google Drawings allowed us to do virtual sessions, thus minimizing the risks for the participants. Finally, it also provided a detailed editing history which, in conjunction with the recorded sessions, was useful to precisely analyze the steps taken by the participants. 

\section{Narrative Map Construction Strategies}
\label{sec:rq1}
We now present our results, including the cognitive connections, the construction strategies, and the graph and layout properties of the maps. Throughout this process, we used open coding to discover the different types of cognitive strategies and analyze the results.

\subsection{How do analysts connect events?}
To answer this question, we asked participants to explain their connection strategies as they constructed the map as well as in the follow-up interviews. We identified seven types of connections, which we further divided into low-level, high-level, and supporting connections. Low-level connections are those that can be made directly from the content of the document (e.g., dates, keywords, entities present) without an in-depth analysis. In contrast, high-level connections involve applying cognitive schemas to synthesize information between events \cite{bradel2013analysts}. Supporting connections are used in conjunction with high-level connections as an auxiliary strategy to help connect events. For example, a connection could be based on cause-effect relationships between events (high-level connection) and speculation (supporting connection). These results show that beyond offering a chronological chain of events, it would be beneficial to visualize narratives based on  different types of relationships, such as those based on similarity, entities, topics, and causality. 

\subsubsection{Low-level Connections}
We identified 3 low-level connections (temporal, similarity, and entity) based on our analysis of the sessions with the participants.

\begin{enumerate}[itemsep=0em, wide, labelindent=0pt]
    \item \textbf{Temporal connections} (A happened before B): Most participants used temporal connections as their default strategy. In particular, this type of connection was used when there were no other explicit relationships between events and participants wanted to maintain the temporal sequence of the events (e.g., ``I just followed chronological order"). All participants used temporal connections in their maps, because of the inherent chronology of narratives.
    \item \textbf{Similarity connections} (A is similar to B): Users determined two events as similar primarily based on keyword matching and a superficial similarity evaluation (e.g., ``All these events mentioned markets or production''). 
    \item \textbf{Entity connections} (A is about the same entity as B): These connections are based on named entity co-occurrences in events. For example, some participants focused on whether the events referred to specific entities (e.g., ``These events talk about China").
\end{enumerate}

\subsubsection{High-level Connections}
We classified connections as high-level if they involved the use of a cognitive schema to connect information between documents. In particular, these connections usually arise from inferences made by the users rather than a superficial characteristic of the document. 

\begin{enumerate}[itemsep=0em, wide, labelindent=0pt]
\item \textbf{Topical connections} (A shares a common theme or topic with B): These connections are a more abstract version of the similarity connection. They focus on the overarching topic or theme of the articles (e.g., ``These events are about the Chinese government response"). They differ from their low-level counterpart because they are based on a semantic viewpoint rather than superficial similarity. 
\item \textbf{Causal connections} (A leads to B): These relationships join events if one is caused (or could be caused) by another (e.g., ``The number of cases surpassing SARS led to stricter travel restrictions"). 
Some causal relationships defy the temporal order because the reporting date is not the actual event date, leading to participants changing the order of the documents to respect the cause-effect relation.
\end{enumerate}

\subsubsection{Supporting Connections}
We classified connections as supporting if they are auxiliary strategies used in conjunction with a high-level connection. 

\begin{enumerate}[wide,itemsep=0em, labelindent=0pt]
\item \textbf{Speculative connections} (A is connected to B because of inferred X): These connections are based on implications based on the participant's beliefs (e.g., ``The call for a global ban on animal markets made the global markets panic", a paraphrased causal speculative connection from one of the participants). Thus, speculative connections relate events that do not share any explicit relationship but could be connected based on speculative reasons. 
\item \textbf{Domain Knowledge connections} (A is related to B because of external knowledge X): These connections are a special type of connection where documents that do not share any explicit relationship are connected based on external domain knowledge (e.g., ``Air travel and oil demand are related").
\end{enumerate}

\subsection{What are analysts' map construction strategies?}
We studied the construction process by following the individual steps taken by the participants as they built their narrative maps. We also asked follow-up questions about the process during the interviews. The identified strategies are summarized in Figure \ref{fig:strats}.

\begin{enumerate}[wide, itemsep=0em, labelindent=0pt]
\item \textbf{Clustering Strategy}: Clustering allows analysts to group documents based on specific characteristics (e.g., topic, type of document, source). Half of the participants had an explicit clustering step during the creation of the map. The use of clustering in sensemaking tasks has also been reported in previous research, either as a story construction strategy \cite{bradel2013analysts} or as the final product \cite{endert2012clustering}. However, in the context of narrative maps, the main purpose of clustering is as a tool to aid in storyline constructions. If the participants performed clustering, it was either done as a preprocessing step or as an intermediate step (i.e., after starting with the connections). We note that clusters are not necessarily the same as storylines, but they can be used as a stepping stone towards identifying the storylines.

\item \textbf{Construction Focus}: This strategy refers to the part of the narrative map that was created first. Participants either focused on the main story, the side stories, or followed no particular order (i.e., a mixed strategy going back and forth). The main story refers to the sequences of core events in the narrative, those that move the narrative forward \cite{keith2020maps, abbott2008cambridge}. In contrast, the side stories do not form part of the narrative core. Instead, they provide further information and useful context to the narrative. Note that there is an even split between focusing on the main story and following a mixed strategy. 

\item \textbf{Algorithm Type}: This strategy refers to the general algorithm that participants followed to construct the map. By analyzing the order in which participants constructed the maps, we found three types of strategies. The first two strategies are conceptually similar to basic graph searching algorithms---constructing the map in a depth-first or breadth-first fashion---while the third strategy is based on clustering---turning clusters into storylines. Note that depth-first approaches either focused on side stories first or on the main story first. In contrast, breadth-first approaches followed a mixed strategy by definition. The strategy of turning clusters into storylines either focused on side stories first or followed a mixed strategy.

\begin{figure}[!htb]
    \centering
    \includegraphics[width=0.85\columnwidth]{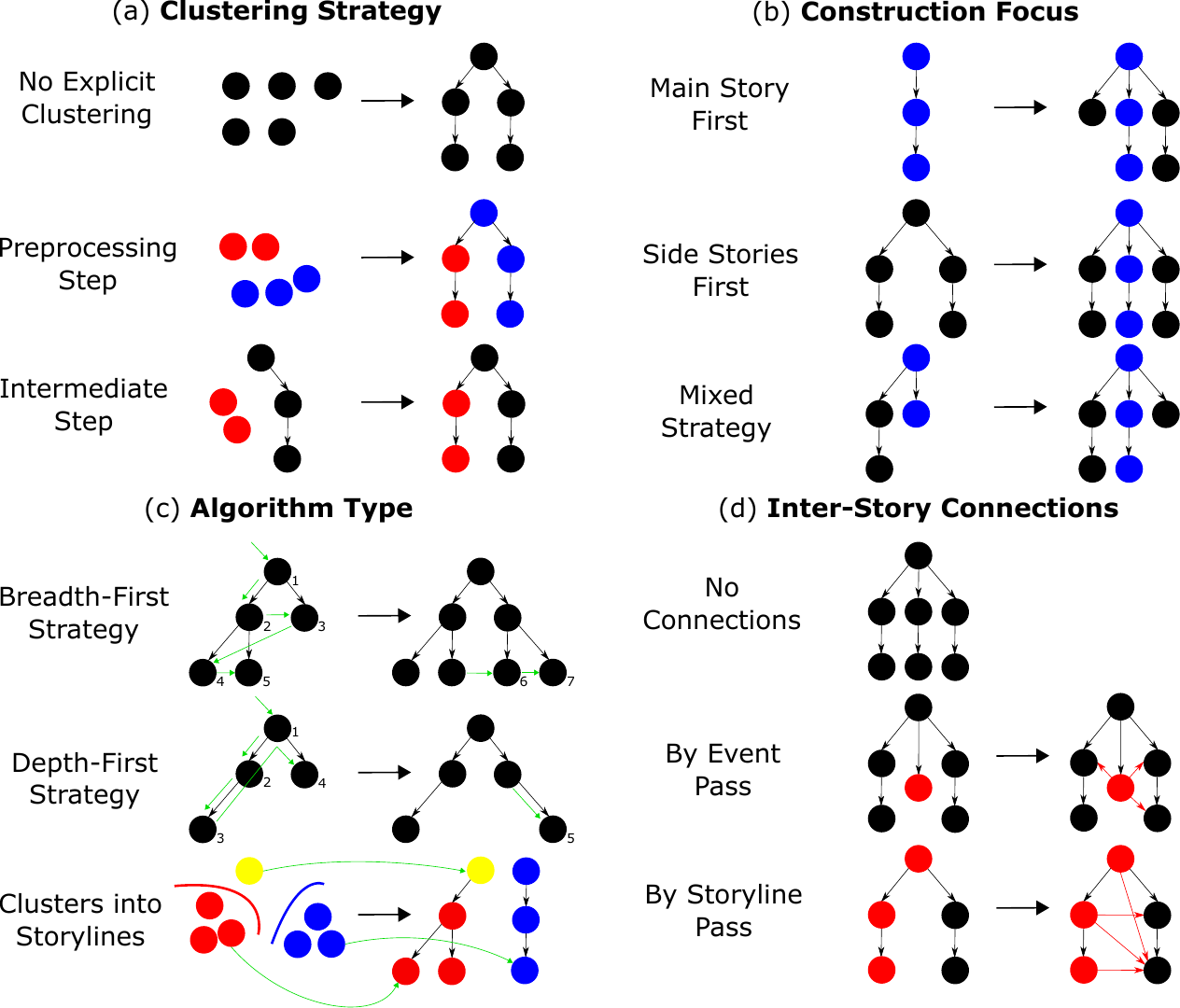}
    \caption{Narrative map construction strategies. (a) \textbf{Clustering strategies}: no clustering, clustering during preprocessing, or clustering in the middle of construction. b) \textbf{Construction focus}: whether participants created the main story (in blue) first, the side stories first, or a mixed strategy. (c) \textbf{Algorithm type}: the order in which nodes were added, following either a breadth-first, depth-first, or clustering approach. (d) \textbf{Inter-story connections}: some participants checked for inter-story connections (in red) when adding events, others checked when completing a storyline.}
    \label{fig:strats}
    \vspace{-5pt}
\end{figure}

\item \textbf{Inter-story Connections}: This strategy refers to how the participants connected storylines. In most cases, participants did not add inter-story connections; making their storylines independent from the rest of the graph, except for the initial connection where they split off. In other cases, they added connections on a by-event basis, checking whether an event should be connected to other stories as they add it. Alternatively, they added connections on a by-storyline basis, checking whether to connect the storyline with others only after completing the whole storyline.
\end{enumerate}

\subsection{What are the properties of the created maps?}
We focus on multiple structural aspects of the underlying graph and the layout considerations made by participants (see Table \ref{tab:layout}).

\begin{table}[!htb]
\centering
\small
\resizebox{0.80\columnwidth}{!}{%
\begin{tabular}{ccccc}
\hline
\textbf{Property}                             & \textbf{Code}                                                                                             & \textbf{DT} & \textbf{OT} & \textbf{Total} \\ \hline
\multirow{3}{*}{\textbf{Graph Structure}}     & List \protect\resizebox{0.35cm}{!}{\protect\includegraphics{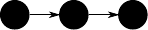}}                             & 1                      & 1                        & 2              \\
                                              & Tree \protect\resizebox{0.30cm}{!}{\protect\includegraphics{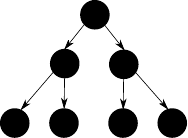}}                             & 2                      & 2                        & 4              \\
                                              & DAG \protect\resizebox{0.20cm}{!}{\protect\includegraphics{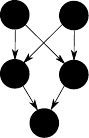}}                               & 2                      & 2                        & 4              \\ \hline
\multirow{3}{*}{\textbf{Layout}}              & Vertical (top-down) \protect\resizebox{0.20cm}{!}{\protect\includegraphics{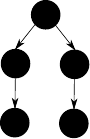}}          & 3                      & 4                        & 7              \\
                                              & Diagonal (left to right) \protect\resizebox{0.30cm}{!}{\protect\includegraphics{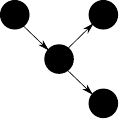}}     & 1                      & 0                        & 1              \\
                                              & Horizontal (left to right) \protect\resizebox{0.30cm}{!}{\protect\includegraphics{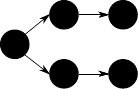}} & 1                      & 1                        & 2              \\ \hline
\multirow{2}{*}{\textbf{Main Story Position}} & Main Story First \protect\resizebox{0.25cm}{!}{\protect\includegraphics{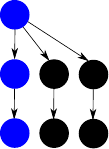}}           & 4                      & 2                        & 6              \\
                                              & Main Story Center \protect\resizebox{0.25cm}{!}{\protect\includegraphics{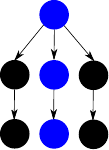}}         & 1                      & 3                        & 4              \\ \hline
\multirow{2}{*}{\textbf{Source Nodes}}        & Single \protect\resizebox{0.20cm}{!}{\protect\includegraphics{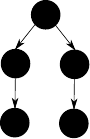}}                   & 4                      & 2                        & 6              \\
                                              & Multiple \protect\resizebox{0.20cm}{!}{\protect\includegraphics{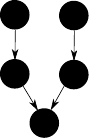}}               & 1                      & 3                        & 4              \\ \hline
\multirow{2}{*}{\textbf{Sink Nodes}}          & Single \protect\resizebox{0.20cm}{!}{\protect\includegraphics{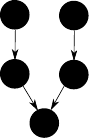}}                     & 2                      & 0                        & 2              \\
                                              & Multiple \protect\resizebox{0.20cm}{!}{\protect\includegraphics{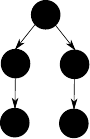}}                 & 3                      & 5                        & 8              \\ \hline
\multirow{2}{*}{\textbf{Connectivity}}        & Connected \protect\resizebox{0.20cm}{!}{\protect\includegraphics{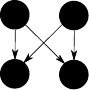}}                   & 4                      & 4                        & 8              \\
                                              & Disconnected \protect\resizebox{0.20cm}{!}{\protect\includegraphics{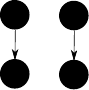}}             & 1                      & 1                        & 2              \\ \hline
\multirow{2}{*}{\textbf{Transitivity}}        & Implicit \protect\resizebox{0.30cm}{!}{\protect\includegraphics{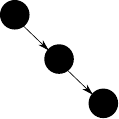}}                     & 5                      & 3                        & 8              \\
                                              & Explicit \protect\resizebox{0.30cm}{!}{\protect\includegraphics{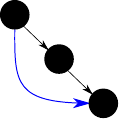}}                 & 0                      & 2                        & 2              \\ \hline
\end{tabular}%
}
\caption{Graph and Layout Properties for each participant in our study.}
\label{tab:layout}
\vspace{-15pt}
\end{table}

\begin{enumerate}[wide, itemsep=0em, labelindent=0pt]
\itemsep0em
\item \textbf{Graph Structure}: We found that participants used three types of underlying graph structures: lists \protect\resizebox{0.35cm}{!}{\protect\includegraphics{graphs/list.pdf}}, trees \protect\resizebox{0.35cm}{!}{\protect\includegraphics{graphs/tree.pdf}}, and directed acyclic graphs (DAGs) \protect\resizebox{0.20cm}{!}{\protect\includegraphics{graphs/dag.pdf}}. These results are in line with prior work on narrative representations, which has focused on similar structures to represent stories \cite{keith2020maps}, such as timelines \cite{shahaf2010connecting}, trees \cite{ansah2019graph}, or other graph variants \cite{yang2009discovering}. Structures were evenly split between trees and DAGs, with only two list-like graphs, where one of them was a single timeline and the other comprised three parallel timelines \protect\resizebox{0.35cm}{!}{\protect\includegraphics{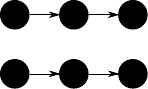}}.

\item \textbf{Layout and Main Story Position}: 
Most participants went for a vertical (top-down) approach \protect\resizebox{0.20cm}{!}{\protect\includegraphics{graphs/vertical.pdf}} with storylines presented as parallel columns and the main story placed first \protect\resizebox{0.25cm}{!}{\protect\includegraphics{graphs/main-first.pdf}} (i.e., the left-most story in a vertical layout or the top story in a horizontal layout). Horizontal layouts \protect\resizebox{0.30cm}{!}{\protect\includegraphics{graphs/horizontal.pdf}} were not preferred; as noted by participants, computer displays seem to favor vertical layouts due to how scrolling works. Finally, one participant used a unique diagonal layout \protect\resizebox{0.30cm}{!}{\protect\includegraphics{graphs/diagonal.pdf}}; we did not observe this behavior in any of the other participants.

\item \textbf{Number of Source and Sink Nodes}: Most people had multiple storyline endings (i.e., sink nodes) \protect\resizebox{0.20cm}{!}{\protect\includegraphics{graphs/multiple-end.pdf}}. In particular, all five open-ended task maps had multiple endings. In contrast, the directed task had two participants constrain themselves to a single ending as defined by the task \protect\resizebox{0.20cm}{!}{\protect\includegraphics{graphs/single-end.pdf}}, while the others added endings or dead-end events for some of the other storylines. For source nodes, participants that had the directed task were more likely to have a single source \protect\resizebox{0.20cm}{!}{\protect\includegraphics{graphs/single-start.pdf}} than those that had the open-ended task \protect\resizebox{0.20cm}{!}{\protect\includegraphics{graphs/multiple-start.pdf}}. The tendency of open-ended maps to have multiple sources and sinks intuitively makes sense given the unrestricted nature of the task. In contrast, the directed task maps are focused on just answering the main question, thus leading to structures that did not have as many loose ends.

\item \textbf{Connectivity}: In graph-theoretical terms, we classify the map as connected if its underlying graph is weakly connected (i.e., we disregard the direction of the arrows). Most participants created connected graphs \protect\resizebox{0.20cm}{!}{\protect\includegraphics{graphs/connected.pdf}}, but there were two cases with separate components \protect\resizebox{0.20cm}{!}{\protect\includegraphics{graphs/disconnected.pdf}}. The first had a separate component for the ``social response and effects of COVID". The second had three parallel timeline structures without any explicit connection between them. 
\item \textbf{Transitivity}: We considered whether participants explicitly included connections that are implied by transitivity (i.e., \protect\resizebox{0.25cm}{!}{\protect\includegraphics{graphs/transitivity.pdf}} vs. \protect\resizebox{0.25cm}{!}{\protect\includegraphics{graphs/nontrans.pdf}}. Most people did not include transitive connections. Only two participants used explicit transitive connections. 
However, even in those cases, they were scarce, meaning that transitive connections were either not needed or participants had difficulty finding them.
\end{enumerate}

\section{Limitations}
Our work is not without limitations. In particular, we conducted interviews with only a handful of participants (10). While the number was small, all participants had a background in intelligence analysis. They also spanned a variety of majors and had reasonable gender representation (6 females and 4 males). Participants had different levels of expertise on the topic and were able to bring insights from their own knowledge and experiences. We note that experience and prior knowledge might influence the strategies, especially with a high-profile topic. Moreover, our data is strictly focused on COVID-19, which may bias participants to adopt specific strategies. However, even with these limitations, we were able to observe diverse construction strategies and map structures. Finally, we note that both of the tasks used in this study represent simplified and constrained versions of what analysts would do in a real-world setup, but they still provide valuable insights into the narrative sensemaking process. 

\section{Conclusions}
\label{sec:conclusions}
We studied how analysts construct narrative maps and the characteristics of these maps. In particular, our user study detected 7 types of cognitive connections. We have shown the importance of topical and causal relationships in the construction of narrative maps, as these were the most common high-level connections in the user-generated maps. In terms of strategies, we found three major ways to construct maps. Each one of these strategies can be the basis of a new narrative extraction algorithm. Furthermore, in terms of the structure of the map, we saw an even distribution between tree-like maps and DAG-like maps. Regarding layout, we found that most users preferred a vertical top-down layout, with the main story shown first. These results can be used to define a series of design guidelines for narrative maps, as well as guide extraction algorithms and interactive visualization tools. Thus, future work should focus on developing specific design guidelines based on these results. Moreover, it would also be useful to explore how strategies differ when applied to different domains, data set sizes, and analyst experience. In particular, it would be useful to consider how previous analyst training (e.g., experience with structured analytic techniques) could influence the construction strategies or the narrative map structures.

\acknowledgments{
We would like to thank the InfoVis Lab at Virginia Tech and the Social Computing Lab at the University of Washington for their valuable comments and feedback on early drafts of the paper. This work was partially funded by NSF grants CNS-1915755 and DMS-1830501 and by ANID/Doctorado Becas Chile/2019 - 72200105.}

\bibliographystyle{abbrv-doi}

\balance
\bibliography{template}
\end{document}